\documentclass[
    ,final            
  ]
  {aipproc}

\layoutstyle{6x9}

\usepackage{amsmath}
\usepackage{amssymb}
\usepackage{amsfonts}

\newcommand{\pibf}{\mbox{\boldmath $\pi$}}

\newcommand{\veceps}{\mbox{\boldmath$\epsilon$}}
\newcommand{\vecsig}{\mbox{\boldmath$\sigma$}}

\begin{document}

\title{Scalar mesons and polarizability of the nucleon }


\author{Martin Schumacher}{
  address={Zweites Physikalisches Institut der Universit\"at G\"ottingen,
Friedrich-Hund-Platz 1\\ D-37077 G\"ottingen, Germany }
}

\begin{abstract}
It is  shown that the scalar mesons $\sigma$, $f_0(980)$ and $a_0(980)$ 
as $t$-channel exchanges quantitatively solve the problem of diamagnetism
and give an  explanation of the large missing part of the
electric polarizability
$\alpha$ showing up when only the pion cloud is taken into account. The
electric polarizability of the proton $\alpha_p$
confirms a two-photon width
of the $\sigma$ meson of $\Gamma_{\sigma\gamma\gamma}= (2.58\pm 0.26) $ keV.
\end{abstract}

\maketitle


\section{Introduction}
Due to recent experiments very precise data are available for the
polarizabilities of the nucleon \cite{schumacher05}.
The interpretation of these data in terms of  the structure of the
nucleon was a problem up to very recently. Calculations using nucleon
models or chiral perturbation theory were not sufficient. Also calculations
in terms of dispersion theory failed, unless contributions from a scalar
$t$-channel were taken into account which, however, were only
vaguely known. Recently essential progress has been made by showing
\cite{schumacher06,schumacher07a,schumacher07b} that the largest part 
of the $t$-channel contributions can be predicted from the 
 $\sigma$ meson having properties as proposed by Delbourgo and Scadron
\cite{delbourgo95}. Minor contributions are from the
$f_0(980)$ and $a_0(980)$ mesons.

\section{Compton scattering and polarizabilities}

\begin{figure}
\includegraphics[width=0.6\linewidth]{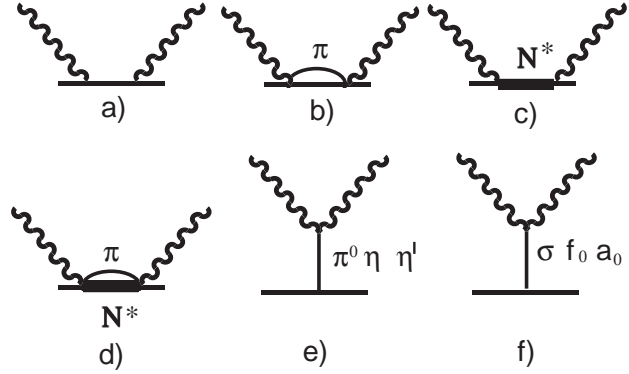}
\caption{Graphs describing Compton scattering by the nucleon. 
Crossed graphs are not shown. Graph a) represents the Born terms.
Graphs b) -- d) have counterparts in the
photoabsorption cross section of the nucleon and, therefore, contribute
to the $s$-channel part of the Compton scattering amplitude. These graphs
represent the contributions of the resonant and nonresonant nucleon excited
states to the polarizabilities.
The graphs e) and f) may be viewed as properties
of the constituent quarks which are coupled to the fields of the respective
particles.  These graphs make contributions to the $t$-channel part
of the Compton scattering amplitude. Pseudoscalar mesons couple to two
photons with perpendicular planes of linear polarization and make
contributions to $\gamma_\pi$. Scalar mesons couple to two photons with 
parallel planes of linear polarization and make contributions to
$(\alpha-\beta)$. }
\label{graphs} 
\end{figure}

The polarizabilities of the nucleon are determined via Compton scattering.
For the forward and backward direction the amplitudes are given by 
(\ref{T0}) and (\ref{Tpi}), respectively,
\begin{eqnarray}
&&\frac{1}{8\pi m}[T_{fi}]_{\theta=0}
=f_0(\omega){\veceps}'{}^*\cdot{\veceps}+
g_0(\omega)\, \mbox{i}\, {\vecsig}\cdot({\veceps}'{}^*\times
{\veceps}), \label{T0}\\
&&\frac{1}{8\pi m}[T_{fi}]_{\theta=\pi}
=f_\pi(\omega){\veceps}'{}^*\cdot{\veceps}+
g_\pi(\omega)\, \mbox{i}\,{\vecsig}\cdot({{\veceps}}'{}^*
\times
{{\veceps}}),
\label{Tpi}
\end{eqnarray}
where $m$ is the nucleon mass, 
$\omega$  the energy of the incoming photon in the LAB system, 
${\vecsig}$  the spin operator of the nucleon and
${\veceps}$ and ${\veceps}'$ the polarization vectors of the incoming and
outgoing photon.
A definition of the electric ($\alpha$) and  magnetic ($\beta$) 
polarizabilities
and the spin polarizabilities $\gamma_0$ and $\gamma_\pi$ for the forward
and backward directions is  obtained by 
expanding the amplitudes in terms of the photon energy $\omega$
 \begin{eqnarray}
f_0(\omega) & = & - ({e^2}/{4 \pi m})q^2 + 
{\omega}^2 ({\alpha}
+{\beta}) + {\cal O}({\omega}^4), \label{f0}\\
g_0( \omega) &=&  \omega\left[ - ({e^2}/{8 \pi m^2})\,   
{\kappa}^2 
+ {\omega}^2
{\gamma_0}  + {\cal O}({\omega}^4) \right], \label{g0}\\
f_\pi(\omega) &=& \left(1+({\omega'\omega}/{m^2})\right)^{1/2} 
[-({e^2}/{4 \pi m})q^2 +       
\omega\omega'({\alpha} - {\beta}) 
+{\cal O}({\omega}^2{{\omega}'}^2)], \label{fpi}\\
g_\pi(\omega) &=& \sqrt{\omega\omega'}[
({e^2}/{8 \pi m^2})  
( {\kappa}^2 + 4q
{\kappa} + 2q^2)
+ \omega\omega'
{{\gamma_\pi}} + {\cal O}
({\omega}^2 {{\omega}'}^2)],  \label{gpi}\\
\omega'&=&\omega/(1+(2\omega)/m), \nonumber
\end{eqnarray}
where $q e$ is the electric charge ($e^2/4\pi=1/137.04$) and
$\kappa$ the anomalous magnetic
moment of the nucleon.
The relevant graphs are shown in Figure \ref{graphs}. Graph a) represents the 
Born terms, b)  the single pion meson-cloud contribution,
c) the contributions of nucleon resonances, d) higher-order
$s$-channel contributions, e) the pseudoscalar $t$-channel and
f) the scalar $t$-channel.

The $s$-channel parts of the electromagnetic polarizabilities can be 
calculated from the multipole components of the photoabsorption cross section
of the nucleon using the BL sum rule (see \cite{schumacher05})
\begin{equation}
(\alpha+\beta)^s=\frac{1}{2\pi^2}
\int^\infty_{\omega_{\rm thr}}\frac{\sigma_{\rm tot}(\omega)}{\omega^2}d\omega
\label{Eq-a}
\end{equation}
for the forward direction and the BEFT sum rule (see \cite{schumacher05})
\begin{equation}
(\alpha-\beta)^s=\frac{1}{2\pi^2}\int^\infty_{\omega_{\rm thr}}
\sqrt{1+\frac{2\omega}{m}}\left(\sigma_{E1,M2,\cdots}(\omega)-
\sigma_{M1,E2,\cdots}(\omega)\right)\frac{d\omega}{\omega^2}
\label{Eq-b}
\end{equation}
for the backward direction. The results are given in Table \ref{schannel}.
\begin{table}[h]
\caption{$s$-channel contributions to the electromagnetic polarizabilities
(see \cite{schumacher07a,schumacher07b})}
\begin{tabular}{llcccc}
&&$\alpha_p$&$\beta_p$& $\alpha_n$&$\beta_n$  \\
\hline
$P_{33}(1232)$&$M1,E2$&$-1.1$&{$+8.3$}& $-1.1$&{$+8.3$}          \\
$P_{11}(1440)$&$M1$&$-0.1$&$+0.3$&  $-0.1$&$+0.3$     \\
$D_{13}(1520)$&$E1,M2$&$+1.2$ &$-0.3$&  $+1.2$ &$-0.3$        \\
$S_{11}(1535)$&$E1$&$+0.1$ &$-0.0$& $+0.1$ &$-0.0$         \\
$F_{15}(1680)$&$E2,M3$&$-0.1$ &$+0.4$& $-0.1$ &$+0.4$          \\
{nonresonant}&&$+4.5$ & $+0.7$ & $+5.6$& $+0.9$
\label{schannel}
\end{tabular}
\end{table}

\section{$t$-channel contributions to the polarizabilities}

In our present approach the $\sigma$ meson
simultaneously leads to mass generation of the 
constituent quarks and to the main part of the $t$-channel contribution
to $\alpha-\beta$. Therefore, 
the polarizabilities provide us with access to the off-shell $\sigma$-meson
as entering into dynamical symmetry breaking.  Dynamical symmetry breaking
starts from  equations for the pion decay constant $f^{\rm cl}_\pi=89.8$
MeV in the chiral limit (cl) and the mass $M$ of the constituent quark in the 
chiral limit \cite{schumacher06,delbourgo95}
\begin{equation}
f^{\rm cl}_\pi=-4 i N_c\, g M \int \frac{d^4p}{(2\pi)^4}
\frac{1}{(p^2-M^2)^2},\,\,
M=-\frac{8iN_c\,g^2}{(m^{\rm cl}_\sigma)^2}\int\frac{d^4p}{(2\pi)^4}
\frac{M}{p^2-M^2},
\label{dynamical}
\end{equation}
where $m^{\rm cl}_\sigma=2M$ is the $\sigma$ meson mass in the chiral limit and
$N_c=3$ the number of colors.
Using dimensional regularization and making use of the Goldberger-Treiman
relation this leads to 
\begin{equation}
g=M/ f^{\rm cl}_\pi  =2\pi/\sqrt{3}=3.63 \quad \mbox{and} \quad M=325.8. 
\label{Eq-c}
\end{equation}
Furthermore, using the current quark masses $m^{\rm curr.}_u=3.0$ MeV
and $m^{\rm curr.}_d=6.5$ MeV we arrive at the constituent quark masses
$m_u=328.8$ MeV, $m_d=332.3$ MeV. The mass of the $\sigma$ meson
is given by  $m_\sigma=(4M^2+\hat{m}^2_\pi)^{1/2}=666.0$ MeV, where 
$\hat{m}_\pi$ is the average pion mass. This value  $m_\sigma=666.0$ MeV
for the sigma meson mass will be used throughout 
in the following calculations.

For pseudoscalar and scalar mesons  having the constituent quark structure
\begin{equation}
|q\bar{q}\rangle=a|u\bar{u}\rangle+b|d\bar{d}\rangle+c|s\bar{s}\rangle,
\quad a^2+b^2+c^2=1,
\label{Eq-d}
\end{equation}
the two-photon decay amplitude may be given in the form
\begin{equation}
M(M\to\gamma\gamma)
=\frac{\alpha_e}{\pi f_\pi}N_c\sqrt{2}\langle\, e^2_q\rangle,\quad
\mbox{with} \quad  \langle  e^2_q \rangle 
= a\, e^2_u +b\, e^2_d +c\,(\hat{m}/m_s)\,e^2_s,
\label{M-formula}
\end{equation}
where $\alpha_e=1/137.04$, $f_\pi=(92.42\pm 0.26)$ MeV the pion decay 
constant, $\hat{m}$ the average constituent mass
of the light quarks and $m_s$ the constituent mass of the strange quark.
Numerically we have $m_s/\hat{m}´\simeq 1.44$ \cite{scadron04}. Using      
(\ref{M-formula}) and adjusting the two-photon widths
\begin{equation}
\Gamma_{M\gamma\gamma}=\frac{m^3_M}{64 \pi}|M(M\to\gamma\gamma)|^2
\label{twophoton}
\end{equation}
to the experimental data \cite{PDG} we arrive at the following 
$|q\bar{q}\rangle$ structures of pseudoscalar and scalar mesons

\vspace{2mm}
\begin{tabular}{lll}
\vspace{2mm}
$|\pi^0\rangle$&$\!\!\!\!\!= \frac{1}{\sqrt{2}}|-u\bar{u}+d\bar{d}\rangle$&
$^1S_0$\\
\vspace{2mm}
$|\eta\rangle$&$\!\!\!\!\!=\frac{1}{\sqrt{2}}(1.04\,|IS\rangle-0.96\,|s\bar{s}
\rangle)$&
$^1S_0$\\\vspace{2mm}
$|\eta'\rangle$&$\!\!\!\!\!=\frac{1}{\sqrt{2}}(0.83\,|IS\rangle+1.15\,|s\bar{s}
\rangle)$&
$^1S_0$\\
\vspace{2mm}
$|IS\rangle$&$\!\!\!\!\!=\frac{1}{\sqrt{2}}(|u\bar{u}\rangle+|d\bar{d}
\rangle$&
\\\vspace{2mm}
$|IV\rangle$&$\!\!\!\!\!=\frac{1}{\sqrt{2}}(-|u\bar{u}\rangle+|d\bar{d}
\rangle$&
\\
\end{tabular}

\begin{tabular}{lll}
\vspace{2mm}
$|\sigma\rangle$&$\!\!\!\!\!=\frac{1}{\sqrt{2}}|u\bar{u}+d\bar{d}
\rangle$&$^3P_0$\\
\vspace{2mm}
$|f_0(980)\rangle$&$\!\!\!\!\!=\frac{1}{\sqrt{2}}(0.52\,|IS\rangle-1.32\,
|s\bar{s}
\rangle)$&$^3P_0$\\\vspace{2mm}
$|a_0(980)\rangle$&$\!\!\!\!\!=\frac{1}{\sqrt{2}}(0.83\,|IV\rangle+1.15
\,|s\bar{s}
\rangle)$&$^3P_0$\\ 
\vspace{2mm}
$|IS\rangle$&$\!\!\!\!\!=\frac{1}{\sqrt{2}}(|u\bar{u}\rangle+|d\bar{d}
\rangle$=$-\frac{1}{\sqrt{2}}(|K^+K^-\rangle+|K^0\bar{K}^0\rangle)$&
\\\vspace{2mm}
$|IV\rangle$&$\!\!\!\!\!=\frac{1}{\sqrt{2}}(-|u\bar{u}\rangle+|d\bar{d}
\rangle$=$\frac{1}{\sqrt{2}}(|K^+K^-\rangle-|K^0\bar{K}^0\rangle)$ &
\\
\end{tabular}\\
In these adjustments we have kept the structures of the $\pi^0$ and $\sigma$
meson unmodified as given above.
The structure proposed for the $a_0(980)$ meson contains the problem of
isospin violation. This problem can be solved by noting that the mesons
$f_0(980)$ and  $a_0(980)$ and the $K\bar{K}$ meson pairs have almost equal
masses:
$m(f_0)=980\pm10$ MeV, 
$m(a_0)=984.8\pm 1.4$ MeV, 
$m(K^+K^-)= 987.4$ MeV, $m(K^0\bar{K}^0)=995.3$ MeV.
This means that in addition to a $|q\bar{q}\rangle$ 
expansion the states $|IS\rangle$
and  $|IV\rangle$ may also be expanded in terms of $|K\bar{K}\rangle$ states
where the latter are located in the continuum. Taking into account the
mass differences between $|K^+K^-\rangle$ and $|K^0\bar{K}^0\rangle$
we note that the isospin is an oscillating and therefore not definite
quantity:
\begin{equation}
|IS,IV\rangle=\frac{1}{\sqrt{2}}(-e^{i\Delta mt}
|K^+K^-\rangle-|K^0\bar{K}^0\rangle)
\label{Eq-e}
\end{equation}
where $\Delta m$ is the mass difference
between the two $|K\bar{K}\rangle$ components. It is important
to note that the pseudoscalar mesons are in $^1S_0$ angular momentum
states whereas the  scalar mesons are in $^3P_0$ angular momentum states. 
For the $\pibf$ and the $\sigma$ mesons this means that the $\pibf$ mesons are
located in the ground state of a confining potential whereas 
the $\sigma$ meson is located in the first excited state. 
This implies  that at least qualitatively confinement provides
a complementary explanation of chiral symmetry breaking.

Table \ref{summary} summarizes the properties of the pseudoscalar and scaler
mesons as obtained from the $|q\bar{q}\rangle$ structures given above.
The decay amplitude is obtained using (\ref{M-formula}) and (\ref{twophoton}).
The calculation of the meson-nucleon coupling constant is described in
\cite{schumacher07b}. The quantity  $\Gamma_{M \gamma\gamma}$ is the  
experimental two-photon width, except for the $\sigma$ meson where the
prediction of dynamical symmetry breaking is given. Within the framework
of this theory 
the only error entering into the result is due to the pion decay constant
$f_\pi$  which is known with a precision of $\sim 0.3\%$. This leads to an
error of $\sim 1\%$ for the predicted 
two-photon width $\Gamma_{\sigma\gamma\gamma}$. An experimental error is
obtained from the electric polarizability of the proton which is known with a
precision of $\sim 5\%$. Using 
$\Delta\Gamma_{\sigma\gamma\gamma}/\Gamma_{\sigma\gamma\gamma}\approx
2\Delta\alpha_p/\alpha_p\approx 10\%$ 
the error  adopted in Table \ref{summary} is obtained.
\begin{table}[h]
\caption{Decay amplitudes and meson-nucleon coupling constants corresponding
  to the meson structures given above}
\begin{tabular}{l|l|l|l}
&$M(M\to \gamma\gamma)$&$g_{M\, NN}$&$\Gamma_{M\gamma\gamma}$\\
&$[10^{-2}$ GeV$^{-1}]$&&[keV] \\
\hline
$\pi^0$&$-2.513\pm 0.007$ &$13.169\pm0.057$&$(7.74\pm0.55)\times10^{-3}$
(see \cite{schumacher06}) \\
$\eta$&$+2.50\pm0.06$&$5.79\pm 0.15$&$0.510\pm0.026$ \cite{PDG}\\
$\eta'$&$+3.13\pm0.05$&$4.63\pm 0.08$&$4.29\pm0.15$ \cite{PDG}\\
$\sigma$&+4.19(*)&$13.169\pm0.057$&$2.58\pm 0.26$(*);
$4.1\pm 0.3$ \cite{pennington06};
$1.6\pm 0.2$ \cite{oller08}\\
$f_0$&$+0.79\pm0.11$&$5.8\pm 0.8$&$0.29^{+0.07}_{-0.09}$\cite{PDG}\\
$a_0$&$-0.79\pm 0.13$&$7.7\pm 1.2$&$0.30\pm0.10$\cite{PDG}\\
\hline
&(*) present analysis
\label{summary}
\end{tabular}   
\end{table}
The $t$-channel pole contributions due to scalar mesons are given by
\begin{equation}
(\alpha-\beta)^t=\frac{g_{\sigma NN}\, M(\sigma\to \gamma\gamma)}{2\pi
  m^2_\sigma}+\frac{g_{f_0 NN}\, M(f_0\to \gamma\gamma)}{2\pi
  m^2_{f_0}}+\frac{g_{a_0 NN}\, M(a_0\to \gamma\gamma)}{2\pi
  m^2_{a_0}}\tau_3
\label{alph-bet}
\end{equation}
and 
\begin{equation}
(\alpha+\beta)^t=0.
\label{alph+bet}
\end{equation}
Using these pole contributions the final results given in Table 
\ref{scalarpolarizabilities} are obtained.
\begin{table}[h]
\caption{Electric $(\alpha)$ and magnetic $(\beta)$ polarizabilities for the 
proton $(p)$ and neutron $(n)$. Line 2 contains
resonant and nonresonant contributions due to the nucleon structure
($s$-channel). 
The $t$-channel 
pole contributions may be viewed as properties of the constituent quarks.}
\begin{tabular}{l|llll}
polarizabilities& $\alpha_p$ & $\beta_p$ & $\alpha_n$ & $\beta_n$\\
\hline
resonant + nonresonant&{$+4.5$}& $+9.4$&{$+5.6$}
& $+9.6$\\
$\sigma$ pole&$+7.6$&$-7.6$&$+7.6$&$-7.6$\\
$f_0$ pole&$+0.3$&$-0.3$&$+0.3$&$-0.3$\\
$a_0$ pole&$-0.4$&$+0.4$&$+0.4$&$-0.4$\\
\hline
sum& $+12.0$ & $+1.9$&$+13.9$&$ +1.3$\\
{average exp.  results}& {$12.0\pm 0.6$}&
{$1.9\mp  0.6$}&
{$13.4 \pm 1.0$}&
{$1.8 \mp 1.0$}\\
\hline
&unit $10^{-4}$ fm$^3$&
\label{scalarpolarizabilities}
\end{tabular}
\end{table}
The pseudoscalar $t$-channel contributions are given by
\begin{equation}
\gamma^{\,\,t}_\pi=\frac{1}{2\pi m}\left[\frac{g_{\pi NN}\, 
M(\pi^0\to \gamma\gamma)}{
  m^2_{\pi^0}}\tau_3+\frac{g_{\eta NN}\, M(\eta\to \gamma\gamma)}{
  m^2_{\eta}}+\frac{g_{\eta' NN}\, M(\eta'\to \gamma\gamma)}{
  m^2_{\eta'}}\right].
\label{gamp}
\end{equation}
Inserting the numerical values given in Table \ref{summary} we arrive at
the spin polarizabilities given in Table \ref{spinpolarizabilities}.
\begin{table}[h]
\caption{Spin polarizabilities for the backward direction. Line 2 contains
resonant and nonresonant contributions due to the nucleon structure
($s$-channel) . 
The $t$-channel pole contributions may be viewed as properties of the constituent quarks.} 
\begin{tabular}{l|ll}
spin polarizabilities& $\gamma^{(p)}_\pi$ & $\gamma^{(n)}_\pi$ \\
\hline
resonant + nonresonant&$+(7.1\pm1.8)$&$+(9.1\pm1.8)$\\
$\pi^0$ pole&$-46.7$& $+46.7$ \\
$\eta$ pole&$+1.2$& 
$+1.2$ \\
 $\eta'$ pole&$+0.4$& 
$+0.4$ \\
\hline
sum&$-38.0$&$+57.4$\\
{average exp.  result}& {$-(38.7\pm 1.8)$}&
{$+(57.6\pm 1.8)$}\\
\hline
& unit $10^{-4}$ fm$^4$&
\label{spinpolarizabilities}
\end{tabular}
\end{table}

\section{Summary and discussion}

In the foregoing we have presented information on the structure of 
pseudoscalar and scalar
mesons as entering into the $t$-channel contributions of the polarizabilities
of the nucleon. As $t$-channel exchanges these mesons are probed far
off-resonance and, therefore, may be described in terms of a $q\bar{q}$
expansions of their possibly rather complicated wave-functions. These   
$q\bar{q}$ expansions make it possible to make predictions of their
decay matrix elements $M(M\to \gamma\gamma)$ including their signs
and of the meson-nucleon coupling constants $g_{M NN}$ as 
entering into the expressions for the $t$-channel exchanges. 
For the $\sigma$ meson dynamical symmetry breaking leads to a precise
value $m_\sigma=666.0$ MeV for the mass which is confirmed through the
agreement between the predicted $t$-channel contribution to $(\alpha-\beta)$
and the experimental data.
For all polarizabilities there is a general excellent agreement 
between the experimental and predicted data, showing that the
polarizabilities are known and understood on a high level of precision.

\begin{theacknowledgments}
 The author is indebted to the organizers of the conference Scadron70
(February 11--16, 2008, Lisbon, Portugal) 
for providing
him with the possibility to present the foregoing results.
\end{theacknowledgments}



\bibliographystyle{aipproc}   

\bibliography{sample}

\IfFileExists{\jobname.bbl}{}
 {\typeout{}
  \typeout{******************************************}
  \typeout{** Please run "bibtex \jobname" to optain}
  \typeout{** the bibliography and then re-run LaTeX}
  \typeout{** twice to fix the references!}
  \typeout{******************************************}
  \typeout{}
 }



\end{document}